\documentclass[
 superscriptaddress,
preprint,
 bibnotes,
 longbibliography
 showkeys,
 amsmath,amssymb,
prl,
twocolumn,
]{revtex4-1}

\usepackage{times}

\usepackage{graphicx}
\usepackage{amsmath}
\usepackage{dcolumn}
\usepackage{bm}
\usepackage{braket} 
\usepackage{bbold}
\usepackage{ulem}
\usepackage{epsfig}
\usepackage{subfigure}
\usepackage{float}
\usepackage{comment}
\usepackage{multirow}
\usepackage{color}
\usepackage{lineno}
\usepackage[textsize=footnotesize,textwidth=0.6in]{todonotes}
\usepackage[
 colorlinks=true,
 urlcolor=blue,
 citecolor=blue,
 linkcolor=blue,
 bookmarks=false,
 pdfstartview={FitH},
]{hyperref}

\newcommand{\beginsupplement}{%
        \setcounter{table}{0}
        \renewcommand{\thetable}{S\arabic{table}}%
        \setcounter{figure}{0}
        \renewcommand{\thefigure}{S\arabic{figure}}%
     }

\begin{document}

\title{Spin-polaron ladder spectrum of the spin-orbit-induced Mott insulator Sr$_2$IrO$_{4}$ probed by scanning tunneling spectroscopy}
\author{Jose M. Guevara}
\affiliation{Leibniz-Institute for Solid State and Materials Research, IFW-Dresden, 01069 Dresden, Germany} 
\author{Zhixiang Sun}

\affiliation{Leibniz-Institute for Solid State and Materials Research, IFW-Dresden, 01069 Dresden, Germany}  
 \affiliation{Center for Joint Quantum Studies and Department of Physics, Tianjin University, 300072 Tianjin, China}  
 
\author{Ekaterina M. P\"arschke}
\affiliation{Leibniz-Institute for Solid State and Materials Research, IFW-Dresden, 01069 Dresden, Germany}  
\affiliation{Department of Physics, University of Alabama at Birmingham, Birmingham, Alabama 35294, USA}  
\author{Steffen Sykora}
\affiliation{Leibniz-Institute for Solid State and Materials Research, IFW-Dresden, 01069 Dresden, Germany}  
\author{Kaustuv Manna}\email{Present address: Max-Planck-Institute for 
Chemical Physics of Solids, 01187 Dresden, Germany}
\affiliation{Leibniz-Institute for Solid State and Materials Research, IFW-
Dresden, 01069 Dresden, Germany}  
\author{Johannes Schoop}
\affiliation{Leibniz-Institute for Solid State and Materials Research, IFW-Dresden, 01069 Dresden, Germany}  
\author{Andrey Maljuk}
\affiliation{Leibniz-Institute for Solid State and Materials Research, IFW-Dresden, 01069 Dresden, Germany}
\author{Sabine Wurmehl}
\affiliation{Leibniz-Institute for Solid State and Materials Research, IFW-Dresden, 01069 Dresden, Germany}
\affiliation{Institute for Solid State Physics, TU Dresden, 01069 Dresden, Germany} 

\author{Jeroen van den Brink}
\affiliation{Leibniz-Institute for Solid State and Materials Research, IFW-Dresden, 01069 Dresden, Germany}
\author{Bernd B\"uchner}
\affiliation{Leibniz-Institute for Solid State and Materials Research, IFW-Dresden, 01069 Dresden, Germany}
\affiliation{Institute for Solid State Physics, TU Dresden, 01069 Dresden, Germany}
\affiliation{Center for Transport and Devices, TU Dresden, 01069 Dresden, Germany}
\author{Christian Hess}
\email{c.hess@ifw-dresden.de}
\affiliation{Leibniz-Institute for Solid State and Materials Research, IFW-Dresden, 01069 Dresden, Germany} 
\affiliation{Center for Transport and Devices, TU Dresden, 01069 Dresden, Germany} 
\date{\today}

\begin{abstract}

The motion of doped electrons or holes in an antiferromagnetic lattice with strong on-site Coulomb interactions touches one of the most fundamental open problems in contemporary condensed matter physics. The doped charge may strongly couple to elementary spin excitations resulting in a dressed quasiparticle which is subject to confinement. This  'spin-polaron' possesses internal degrees of freedom with a characteristic 'ladder' excitation spectrum. Despite its fundamental importance for understanding high-temperature superconductivity, clear experimental spectroscopic signatures of these internal degrees of freedom are scarce. Here we present scanning tunneling spectroscopy results of the spin-orbit-induced Mott insulator Sr$_2$IrO$_{4}$. Our spectroscopy data reveal distinct shoulder-like features for occupied and unoccupied states beyond a measured Mott gap of $\Delta\approx620$~meV. Using the self-consistent Born approximation we assign the anomalies in the unoccupied states to the spin-polaronic ladder spectrum with excellent quantitative agreement and estimate the Coulomb repulsion $U$ = 2.05 ...2.28 eV in this material. These results confirm the strongly correlated electronic structure of this compound and underpin the previously conjectured paradigm of emergent unconventional superconductivity in doped Sr$_2$IrO$_{4}$.

\end{abstract}



\maketitle

\begin{figure*}[!t]
\begin{center}
\includegraphics{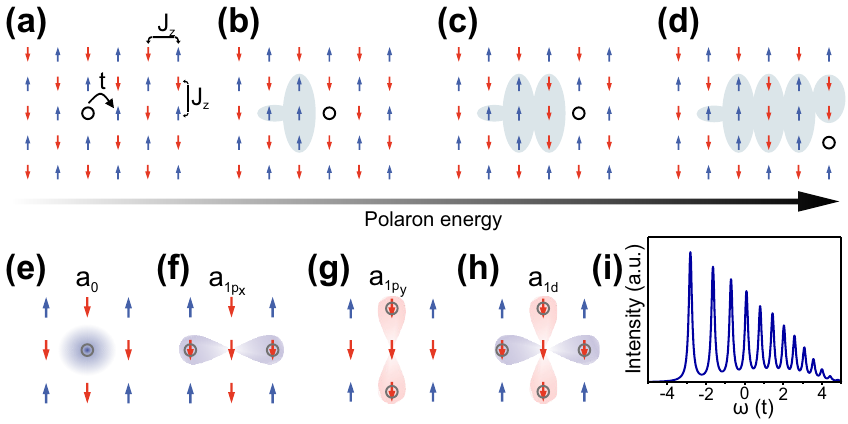}
\end{center}
 \caption{\label{Fig:spinpol1}Illustration of the nature of the spin polaron. Its simplest form is captured by the so-called \textit{t-$J_z$} model which describes a hole in an antiferromagnetic background of Ising spins with $t$ the kinetic energy of the hole and $J_z$ the antiferromagnetic Ising exchange energy between neighboring spins. Panels (a-d) illustrate that a gain in kinetic energy through hopping processes of the hole immediately results in increasing magnetic energy through the destruction of the antiferromagnetic correlation. Thus, the motion of the hole is inhibited and it remains \textit{confined} to its original position.
 Nevertheless, the hole in interaction with the antiferromagnetic background possesses well-defined excited states with different orbital character, the wavefuctions of which (Reiter functions) resemble atomic states. These are characterized by the size (in terms of the number of virtual hopping processes) of the spin polaron, the wavefunctions' symmetry, and a discrete \textit{ladder} excitation spectrum. (e) $s$-wave like ground state of the spin polaron. (f-h) First excited states corresponding to $p$-wave (f,g) and $d$-wave-like states (h).   $a_{0}$, $a_{1 p_{x}}$, $a_{1 p_{y}}$, $a_{1 d}$ denote the coefficients of the Reiter wave function. (i) Ladder spectrum of the \textit{t-$J_z$} model \cite{Martinez1991} where the exchange interaction  $J_z =0.2t$  lies in the parameter region relevant for Sr$_2$IrO$_4$. See Refs.~\onlinecite{Kane1989,Martinez1991,Reiter1994}
  and the Supplementary Materials (SM) for details.  }
\end{figure*}

The so-called spin polaron, describes the motion of a single charge (hole or doublon) added to an antiferromagnetic and insulating ground state of an effective correlated background medium. Thereby, the magnetic excitations of the antiferromagnetic (AF) background can be theoretically  described by a system of bosons (magnons) which couple to the introduced charge carrier via creating virtual bosonic fluctuations. In this way the charge interacts strongly with its environment of ordered spins and forms a new quasiparticle -- the spin polaron. Its excitations have been investigated by, e.g., the self-consistent Born approximation (SCBA) \cite{Martinez1991,Kane1989}, quantum wave function methods \cite{Reiter1994} for a single hole within the $t$-$J$ model, and exact diagonalization \cite{Hamad2008}. These studies show that the spin polaron is characterized by an environment of misaligned spins (Fig.~\ref{Fig:spinpol1}(a-d)) forming an effective confinement potential, where the charge can occupy excited states of different orbital character ~\cite{Wrobel2008} (see Fig.~\ref{Fig:spinpol1}(e-h)). In the one-particle spectral function these excitations manifest themselves by the occurrence of a rather flat and ladder-like structure (Fig.~\ref{Fig:spinpol1}(i)). Due to quantum fluctuations the spin defects can relax and the quasiparticle becomes dispersive.
Despite of the proven fundamental importance for rationalizing many open problems in the physics of correlated electron systems, a direct measurement of the internal degrees of freedom of the spin polaron which proves its peculiar confined nature is still lacking. Here we use scanning tunneling spectroscopy (STS) to specifically probe the excited states of the spin
polaron in a correlated material. Thereby the employed tunneling current serves as to introduce an extra charge (hole or electron, depending on the polarity) into the antiferromagnetic background. We compare the tunneling spectra with theoretical calculations based on the self-consistent Born approximation for the spin-polaronic ladder spectrum and find excellent agreement.

A hallmark of the spin polaron is its connection to the nature of unconventional superconductivity in the cuprates which is believed to emerge from a quasi two-dimensional correlated Mott-insulating antiferromagnetic parent state upon charge doping \cite{Martinez1991}.
The spin polaron and its itinerancy straightforwardly explains the rapid destruction of the antiferromagnetic parent state of the cuprates upon hole doping. Furthermore, it is one key ingredient in many theoretical models which address superconductivity as well as competing phases such as stripe correlations in the underdoped regime of the cuprates' phase diagram \cite{Chernyshev99}.

In recent years, it has been increasingly noticed that quasi-2D iridium oxides exhibit correlated physics that is quite similar to that of the cuprates.
In particular, Sr$_2$IrO$_4$ shares many parallels with the isostructural La$_2$CuO$_4$, a prominent Mott-insulating parent compound of the cuprate high-temperature superconductors \cite{damascelli2003angle}.
The intricate interplay of strong spin-orbit coupling (SOC) and Coulomb repulsion causes the 5$d$ electrons to localize in a state with $J_\mathrm{eff}=1/2$ pseudospins forming the lower Hubbard band of the material with a strong AF exchange interaction \cite{kim2008novel,kim2012magnetic}. AF order occurs below 240 K ($T_N$) \cite{cao1998weak}, and quasi-2D magnon excitations have been detected \cite{kim2012magnetic,Steckel2016}.
In view of these similarities it is reasonable to expect spin polaron physics to be relevant in Sr$_2$IrO$_4$ \cite{Paerschke2017} and it has been argued that a proper doping scheme can drive the material into a high-temperature superconducting phase \cite{wang2011twisted}.



\begin{figure}[!t]
\begin{center}
\includegraphics[width=\columnwidth]{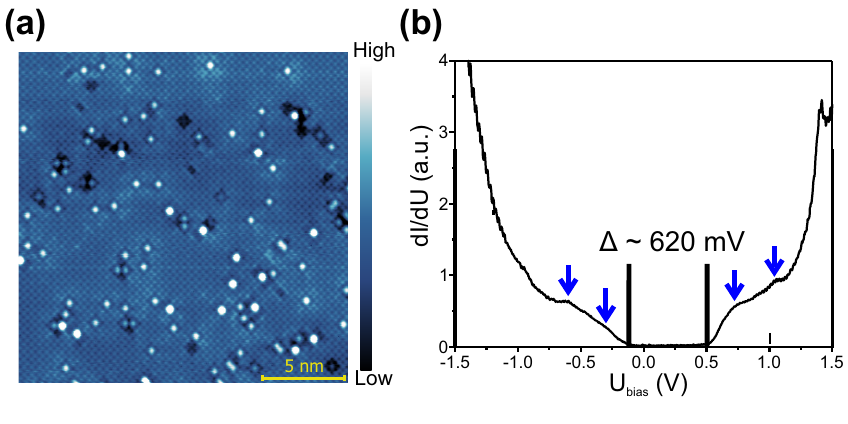}
\end{center}
 \caption{\label{Fig: topo}Topography and the tunneling conductance of the clean area. (a) Topography of the sample surface measured at $T=8.8~$K, with bias voltange $U_\mathrm{bias}=1.0~$V and tunneling current $I_{T}=200~$pA.  (b)  Representative large scale tunneling conductance spectrum taken on a clean place at $T=8.8~$K, with the tip stabilized at $U_\mathrm{bias}=-900~$mV and $I_{T}=200~$pA. Blue arrows indicate the coarse position of fine structure peaks. }
\end{figure}

Low-temperature STM/STS on stoichiometric Sr$_2$IrO$_4$ at $T<10$~K, enabling enhanced spectroscopic resolution, is challenging because samples become too insulating at cryogenic temperatures \cite{dai2014local, Nichols2014, yan2015electron, chen2015influence, Battisti2017, Battisti18}. We therefore used as-grown single crystals of Sr$_2$IrO$_{4-\delta}$ for which a reduced resistivity as compared to the stoichiometric parent compound allows for high-resolution STS measurements even at very low temperature ($T<10$~K) for the first time. Note that the sample is still close to the Mott insulator regime and far away from metallicity, which occurs for high oxygen deficiency \cite{korneta2010electron}, because the resistivity shows a semiconductor-like temperature dependence (see Supplementary Material (SM) Fig. S1). STM data obtained at the cold-cleaved (about 10~K) crystals' surface (Fig.~\ref{Fig: topo}(a)) yield atomically resolved SrO-terminated flat terraces with several local defects (about 2\% with respect to Ir).

Fig. ~\ref{Fig: topo} (b) shows a representative tunneling conductance ($dI/dU$) spectrum taken at 8.8~K at a place free of defects\cite{note2}. The data reveal a sizeable gap $\Delta\approx620$~meV where $dI/dU\approx0$ between about $-110$~mV and $510$~mV,Which we identify as the signature of the Mott gap of the material in agreement with earlier findings  for stoichiometric samples at elevated temperatures \,\cite{dai2014local,yan2015electron}. At bias voltages below and above this gap, the tunneling conductance reveals a shoulder-like increase which straightforwardly can be associated with the density of states of the lower and the upper Hubbard band. At $|U_\mathrm{bias}|\gtrsim 1.3$~V, the $dI/dU$ increases sharply, which we attribute to further high-energy states, in qualitative agreement with optical spectroscopy \cite{propper2016optical}, and a possible energy dependence of the tunneling matrix element. We therefore restrict the following discussion to $U_\mathrm{bias}\lesssim 1.2$~V.

Remarkably, the closer inspection of the $dI/dU$ reveals a distinct fine-structure of peak- or shoulder-like anomalies
(indicated by arrows in Fig.~\ref{Fig: topo} (b)) corroborating earlier studies where already signatures of the first peak at positive bias has been reported \cite{chen2015influence}.  These interesting features indicate a direct coupling of the tunneling electrons/holes to specific excitations of the Mott state. They are ubiquitously present in all spectra taken at  clean areas of the surface (see SM). Indeed, the theoretical investigation of spin polaron physics of Sr$_2$IrO$_4$, which we discuss in the next section, brings into light that this signature is directly related to the excited states of the confined spin polaron quasiparticles including their inherent ladder spectrum.

\begin{figure*}[!t]
\begin{center}
\includegraphics{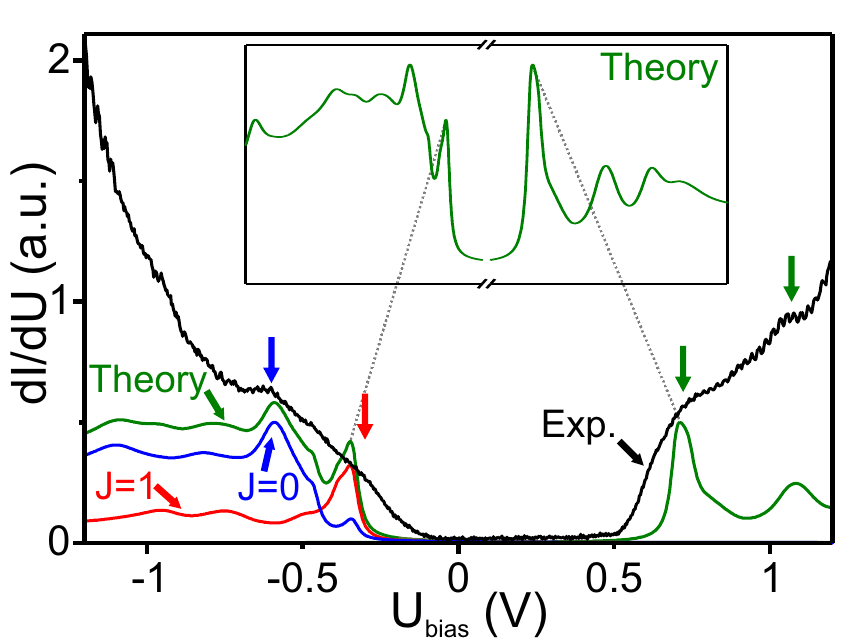}
\end{center}
 \caption{\label{Fig: comp} Comparison of the theoretical and the experimental differential conductance. The green solid line shows the theoretical spectrum calculated from Eq.~\eqref{SW} using the momentum-integrated Green's function of the hole/electron. The black line reproduces the experimental spectrum shown in Fig.~\ref{Fig: topo}(b). At negative bias, the total spectrum is composed of separate contributions due to the multiplet structure of the polaron, i.e. the singlet $J=0$ (blue line) and the triplet $J=1$ (red line) contributions. Since the SCBA inherently provides only the distance between excited levels, the calculated results at positive and negative bias are plotted as to match the respective lowest energy  anomaly in the experimental data. Furthermore, the experimental spectrum has been scaled along the $dI/dU$ axis in order to achieve a rough match with the calculated results. Arrows indicate the coarse position of fine structure peaks as in  Fig.~\ref{Fig: topo} (b). On the negative side equidistant features are not present due to dispersive internal degrees of freedom of the charge excitation. Nevertheless, clear spectroscopic features remain discernible. Inset: Calculated total spectrum from SCBA in a wider energy range.}
\end{figure*}

The clear signatures of the AF Mott physics in the quasi-2D  Sr$_2$IrO$_4$ ~\cite{kim2012magnetic, kim2014excitonic}  suggest that the underlying electron system can be modeled by a multi-orbital 2D Hubbard model with spin-orbit coupling which has an AF ground state as it is known from the usual one-band Hubbard model  ~\cite{Kane1989}. This property simplifies the theoretical treatment by the possibility to apply the well-known	\textquotedblleft 
 single-hole problem\textquotedblright to describe the relevant excitations of the magnetically ordered pseudospin ground state \cite{Schmitt1988,Ramsak1993}. 

Constructing a polaronic model to calculate $dI/dU$ spectra, we address separately the positive and negative bias regions since the strong on-site correlations
render these two cases very different~\cite{Paerschke2017}.

When a negative bias voltage is applied, the electrons are removed from the sample, tunneling towards the tip. This creates an excitation in the
$d^5$ configuration, which can be locally described as a $d^4$ configuration with its complicated intrinsic multiplet structure~\cite{Chaloupka2016}. 
In the lowest energy subspace of the Hilbert space this charge excitation would form a singlet and a triplet state. 
Therefore, to describe a charge excitation on the negative bias side, we introduce a charge excitation creation operator $\textbf{h}$ with an additional degree of freedom
$ |\textbf{h} \rangle   \equiv \{ |J=0 \rangle, |J=1, J_z=1 \rangle, |J=1, J_z=0\rangle, J=1, J_z=-1\rangle \}$. 
Opposite to this, 
applying positive bias voltage, results in adding an electron to the local Ir site with $d^5$ configuration. Hence, 
the charge excitation that one has to consider would resemble the filled-shell $d^6$ configuration and can be described by polaronic excitations shown in Fig.~\ref{Fig:spinpol1}.

The motion of the charge excitation on positive(+) and negative(-) sides of the $dI/dU$ spectra is described by the Hamiltonian:
\begin{linenomath*}
\begin{align}
	    \label{Ham_full_STS}
{\mathcal{H}}^{+,-}={\mathcal{H}}_{\rm mag}+{\mathcal{H}}_{t}^{+,-}, 
\end{align}
\end{linenomath*}
where ${\mathcal{H}}_{\rm mag}$ is the part which includes the low energy excitations of the AF $J_{\mathrm{eff}}=1/2$ ground state. The hopping part of the Hamiltonian, ${\mathcal{H}}_{t}^{+,-}$, 
describes the kinetic energy of the charge coupled to the magnons, which gives rise to the polaron quasiparticle. 
The low-energy effective polaron model
described here has the same operator structure as the effective polaron model of the \textit{t-J} model but has more components than the latter due to the multiplet structure of the polaron. Interestingly,
this additional degree of freedom also allows for more hopping channels, including free (i.~e.~not coupled to magnons) hopping between first neighbors, see SM for details.
The Hamiltonian and its parametrization used here also gives very good quantitative description of the measured ARPES spectra on Sr$_2$IrO$_4$ \cite{Paerschke2017}. Specifically, we have evaluated the one-particle Green's function $G ({\bf k},\omega)$ within the self-consistent Born approximation. 

To relate the described modeling to our measurements we exploit the usual proportionality between the tunneling differential conductance $dI/dU$ and the density of states and calculate the $dI/dU$ using the relation 
\begin{align}
	    \label{SW}
\frac{dI}{dU}(\omega)\backsim -\frac{1}{2\pi} \sum\limits_{{\bf k}} {\rm Im}G ({\bf k},\omega),
\end{align}
where the time evolution in $G ({\bf k},\omega)$ is determined by the  Hamiltonian \eqref{Ham_full_STS}. Fig.~\ref{Fig: comp} shows the results in direct comparison with the experimental data.
In the inset of Fig.~\ref{Fig: comp}, one clearly sees that the calculated spectra on the positive bias side possess ladder spectral features which are similar to those of the much more simplistic spin polaron in terms of the \textit{t-}$J_z$ model (see Fig.~\ref{Fig:spinpol1}(i)) in a remarkable way. 
Indeed, the ladder structure on the positive bias side can be 
clearly identified in the experimental data by the shoulder-like anomalies at  $730 \pm 100$ meV and $970 \pm 100$ meV (see SI for more information about the determination of the position of the structures). We point out that our calculations do not include free parameters. Thus, the almost perfect match between experiment and theory concerning the distance between the ladder peaks corroborates our assignment. Note further, that the experimental tunneling spectra are subject of significant broadening, presumably by electron-phonon scattering processes and higher-order tunneling processes, which account for the differences between the theoretical and the experimental results. 

Not surprisingly however, such a ladder spectrum is not present on the negative bias side -- the polaron motion on the 
negative voltage is additionally greatly complicated by the internal degrees of freedom of the charge excitation, which not only creates additional interacting channels but also
provides a possibility for a nearest-neighbor free hopping of the polaronic quasiparticle.  Therefore, the polaron quasiparticle becomes  more dispersive and a considerable amount of spectral weight is transferred to the incoherent part of the one-particle spectrum. Altogether these two effects lead in the momentum summation in Eq.~\eqref{SW} to a more complex $dI/dU$ on the negative bias side (black line in Fig.~\ref{Fig: comp}). 
Nevertheless, one can study the different contributions to the total Green's function which are carried by  spin polarons of the two different values of the  total quantum momentum, $J=0$ and $J=1$. The calculated contributions are  shown in Fig.~\ref{Fig: comp} in blue and red, correspondingly. Apparently, unlike on the positive bias side, the most salient spectral features correspond separately to singlet and triplet polarons. More specifically, the lowest energy sharp peak is of singlet character whereas the peak at higher energies arises from the triplet polaron. 
Accordingly,  in the experimental data we assign the shoulder at about -300 meV a primary singlet and the peak at about -600 meV a primary triplet character.

After having assigned these most salient aspects of the tunneling spectrum of Sr$_2$IrO$_4$ to essential spin polaron physics, we finally mention that
a careful analysis of the experimental and theoretical spectra shown on Figs. ~\ref{Fig: topo} (b) and ~\ref{Fig: comp} also allows to extract the value of Coulomb repulsion $U$ \cite{note1}.
 It is connected to the Mott gap value $\Delta^{\mathrm{Mott}}$ as
\begin{align}
\label{CoulombMott}
\Delta^{\mathrm{Mott}} = U - E^{\mathrm{pol}}_{\mathrm{hole}} - E^{\mathrm{pol}}_{\mathrm{electron}},
\end{align}
where $E^{\mathrm{pol}}_{\mathrm{hole}}$ ($E^{\mathrm{pol}}_{\mathrm{electron}}$) is the binding energy of the polaron formed when a hole (electron) is added to the ground state of the system.
We estimate polaron binding energies by performing SCBA calculations setting the hopping part of Hamiltonian Eq.~(\ref{Ham_full_STS}) to zero separately for positive and negative bias cases. In this way the polaron
is artificially fully localized and its spectral function is simply a delta function. The binding energies are then given by a relative shift between these delta function peaks and the
quasiparticle peaks of the full calculation (Fig.~\ref{Fig: comp}). From such a consideration the particular  energy values are estimated to be 

\begin{align}
\label{bindingEn}
 E^{\mathrm{pol}}_{\mathrm{hole}} = 0.57\,\mathrm{eV}, \; \; \;     E^{\mathrm{pol}}_{\mathrm{electron}} = 0.81\,\mathrm{eV},
\end{align}

Then the Coulomb repulsion $U$ takes a value between $2.05\,\,\mathrm{eV}$ and $2.18\,\,\mathrm{eV}$ since the Mott gap $\Delta^{\mathrm{Mott}}\approx\Delta$ correct to the lowest quasiparticle peak bandwidth (both
on positive and negative bias sides).

Our findings experimentally and theoretically confirm the important role of the spin-polaronic quasiparticle for the physics of Sr$_2$IrO$_4$. 
More specifically, our data for positive bias voltage, which correspond to electron doping of the AF, reveal clear-cut signatures for the prototypical spin polaron ladder spectrum,
i.e., the fingerprints of the internal degrees of freedom of the electron being confined within the AF background. Thus, in the very low electron doping regime our study reveals yet another similarity between Sr$_2$IrO$_4$ and La$_2$CuO$_4$ concerning  electronic correlations and the inherent consequences, in particular, a possible proximity to superconductivity. More specifically, one may expect that the electron-doped regime of Sr$_2$IrO$_4$ at higher doping levels \textendash\, in analogy to the hole-doped cuprates \cite{wang2011twisted} \textendash\, is very promising to exhibit similar phenomena as the hole doped cuprates. Indeed, the reported signatures of a $d$-wave gap \cite{kim2016observation, yan2015electron} and of stripe-like correlations \cite{Battisti2017} indicate a phenomenology that can be traced back to the spin polaron physics \cite{Chernyshev99, kim2014excitonic}. Thus, the current experimental and theoretical efforts ~\cite{ wang2011twisted, kim2012magnetic, de2015collapse, yan2015electron, kim2016observation, Battisti2017} to find the route to unconventional superconductivity in this material are strongly supported by our study. 
The situation is, however, less clean for the hole-doped regime where the physics is more complicated. Since here the spectral features are dominated by both singlet and triplet polarons, with singlet states being of lower energy with respect to the Fermi level, the analogy to the cuprates is not present. Nevertheless, if chemically achievable, an intricate and fascinating doping evolution governed by the interplay of the singlet and triplet polarons may be expected in the hole-doped regime, too. Finally, we point out that the ubiquitousness of spin polaron physics in all families of Mott insulators makes a pertinent investigation of  high energy spectra of other iridiate \cite{lupke15} and cuprate systems \cite{ye2013}  particularly interesting.

\begin{acknowledgments}
We thank K. Wohlfeld, C. Renner, B.J. Kim and M. Allan for helpful discussions and D. Baumann and U. Nitzsche for technical assistance.
The project is supported by the Deutsche Forschungsgemeinschaft through  SFB 1143 (projects C05, C07, A03, and B01) and by the Emmy Noether programme (S.W. project WU595/3-3). Furthermore, this project has received funding from the European Research Council (ERC) under the European Unions' Horizon 2020 research and innovation programme (grant agreement No 647276 -- MARS -- ERC-2014-CoG).  
\end{acknowledgments}


\makeatletter
\normalem

\bibliography{ir214_spin_polaron_14.bib}

\begin{thebibliography}{33}
\bibitem[{not({\natexlab{a}})}]{note2}
\bibinfo{note}{We constrain our investigation to study defect free areas since
  the AF environment can be expected to be intact, which is not the case at
  defect positions \cite{chen2015influence, Battisti2017}.}
\end{thebibliography}
\bibliographystyle{apsrev}





\newpage
\beginsupplement
\onecolumngrid
\section{Supplementary Materials}

\subsection{Sample preparation}

SrCO$_{3}$ and IrO$_{2}$ powders (with 4N purity) were taken in  stoichiometric ratio and mixed with SrCl$_{2}$ flux with a 1:5 sample-to-flux weight ratio. The mixture was heated to 1210~$^\circ$C for 12~h and then slowly cooled to 1000~$^\circ$C with a cooling rate of 4~$^\circ$C/h, followed by a rapid cool-down to room temperature at 150~$^\circ$C/h. Crystals of Sr$_2$IrO$_{4-\delta}$ with diameter up to 5 mm were filtered from the residue after dissolving the flux in hot water. The crystals were grown in a Pt crucible (50 ccm) with a lid to reduce a flux evaporation.
The crystals  were characterized regarding microstructure (scanning electron microscopy), composition (energy dispersive x-ray analysis), crystallographic structure (single crystal diffraction), magnetic properties (magnetometry), and resistivity.

\subsection{Resistivity data of the Sr$_2$IrO$_{4}$ samples}

\begin{figure}[!t]
\begin{center}
\includegraphics[width=\columnwidth]{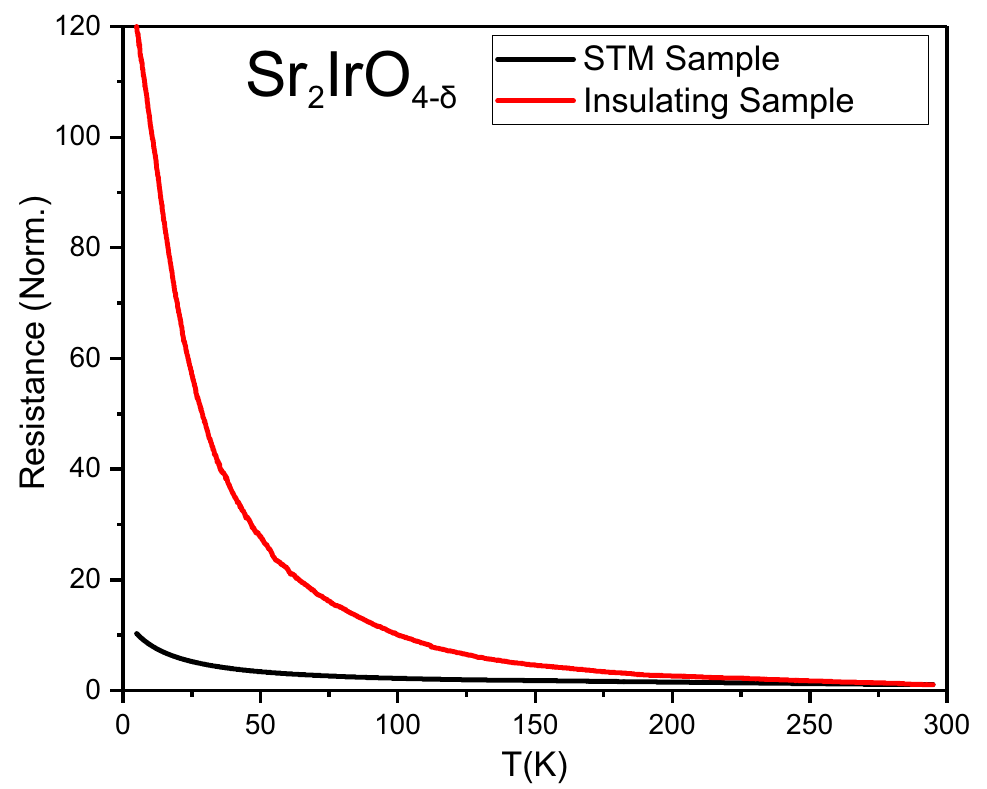}
\end{center}
 \caption{\label{Fig:S1} Normalized in-plane resistivity of selected Sr$_2$IrO$_{4-\delta}$ samples. The strong reduction of the low-temperature upturn of the resistivity of the sample labelled 'STM Sample' evidences a significant amount of oxygen vacancies.}
\end{figure}

The in-plane resistance of the as grown Sr$_2$IrO$_{4-\delta}$ single crystals was measured using a standard 4-probe technique (5 K - 300 K).
Here, $\delta$ accounts for a possible oxygen deficiency.
In the as-grown single crystals, $\delta$ is not controlled which is known to lead to a variation of the resistivity \cite{korneta2010electron}. Fig.~\ref{Fig:S1} shows representative resistivity data for samples with very different temperature dependencies of the resistivity, where the more insulating characterisitics (labelled 'Insulating Sample' in Fig.~\ref{Fig:S1}) can be attributed to an almost stoichiometric oxygen content ($\delta\approx0$) of the corresponding sample \cite{de2015collapse}. On the other hand, the only weakly insulating character of the other sample implies a small amount of oxygen vacancies  \cite{korneta2010electron}. In order to be able to perform high-resolution tunneling spectroscopy at low temperature, we took advantage of the reduced resistivity of this sample (labelled 'STM Sample' in Fig.~\ref{Fig:S1}) and used it for the tunneling experiments in our study. Note that the observed impurity amount of about 2\% with respect to Ir in the topographic data shown in Fig.2(A) is consistent with an oxygen deficiency $\delta$ of the same order \cite{korneta2010electron}.

\subsection{Hamiltonian of the motion of the charge excitation.}
The motion of the charge excitation on positive(+) and negative(-) sides of the $dI/dU$ spectra is described by the Hamiltonian:
\begin{linenomath*}
\begin{align}
	    \label{Ham_full_STS_methods}
{\mathcal{H}}^{+,-}={\mathcal{H}}_{\rm mag}+{\mathcal{H}}_{t}^{+,-}, 
\end{align}
\end{linenomath*}
where ${\mathcal{H}}_{\rm mag}$ describes low energy excitations of the AF $J_{eff}=1/2$ ground state. It is given by
\begin{linenomath*}
 \begin{align}
	    \label{HamHeisenberg}
{\mathcal{H}}_{\rm mag}
=
\sum\limits_{{\bf k}} \omega_{\bf k} (\alpha^\dag_{\bf k} \alpha_{\bf k} + \beta^\dag_{\bf k} \beta_{\bf k}),
\end{align}
\end{linenomath*}
where $\omega_{\bf k}$ is the dispersion of the (iso)magnons represented by the quasiparticle states $| \alpha_{\bf k} \rangle$ and $ | \beta_{\bf k} \rangle $. The hopping part of the Hamiltonian, ${\mathcal{H}}_{t}^{+,-}$, 
describes the transfer of the charge excitation in the bulk coupled to the magnons, which we will also address as polaron quasiparticle. It is given by:
\begin{linenomath*}
\begin{flalign}
	    \label{Hamdparts1}
	    {\mathcal{H}}_{t}^{+} = \sum\limits_{{\bf k}}{V^{0}_{{\bf k}} \left(d^{\dagger}_{{\bf k} A}d_{{\bf k} A} + d^{\dagger}_{{\bf k} B}d_{{\bf k} B}\right)}+\nonumber\\
	     \sum\limits_{{\bf k},{\bf q}} V_{{\bf k},{\bf q}}
	    \left(d^{\dagger}_{{\bf k-q} B}d_{{\bf k} A} \alpha_{\bf q}^{\dagger}+d^{\dagger}_{{\bf k-q} A}d_{{\bf k} B} \beta_{\bf q}^{\dagger} +h.c. \right),
	    \end{flalign}
\end{linenomath*}
	    
\begin{linenomath*}
\begin{align}	    
\label{Hamdparts2}
	   {\mathcal{H}}^{-}_{t}\!= \! \sum\limits_{{\bf k}} \left( {\bf h}_{{\bf k} A}^{\dagger}\hat{V}^{0}_{{\bf k}} {\bf h}_{{\bf k} A}\! +\!{\bf h}_{{\bf k} B}^{\dagger}\hat{V}^{0}_{\bf k} {\bf h}_{{\bf k} B} \right)\! +\! \nonumber\\
	    \sum\limits_{{\bf k}, {\bf q}} \left(  {\bf h}_{{\bf k-q} B}^{\dagger} \hat{V}^{\alpha}_{{\bf k},{\bf q}} {\bf h}_{{\bf k} B} \alpha_{\bf q}^{\dagger}  \!+\!
  {\bf h}_{{\bf k-q} A}^{\dagger}  \hat{V}^{\beta}_{{\bf k},{\bf q}} {\bf h}_{{\bf k} B} \beta_{\bf q}^{\dagger} \!+\! h.c. \right)\!,	    
\end{align}
\end{linenomath*}

Where $A, B$ are the two AF sublattices. The dispersions $\propto V^{0}_{{\bf k}}$, ( $\propto \hat{V}^{0}_{{\bf k}}$) describe the nearest, next nearest, and third neighbor free hopping. 
The terms $\propto V_{{\bf k},{\bf q}}$ ($\propto \hat{V}^{\alpha}_{{\bf k},{\bf q}}$
and  $\propto \hat{V}^{\beta}_{{\bf k},{\bf q}}$) are vertices describing the polaronic hopping of the charge excitation on the positive (negative) side and are given explicitly (also see Ref.~\cite{Paerschke2017}). All the vertices were obtained analytically in a limit of strong on-site Coulomb repulsion and depend on the five hopping parameters of
the minimal tight-binding model obtained as the best fit of the latter to the LDA calculations. The model used here is based on the polaronic model we used to calculate ARPES spectra on Sr$_2$IrO$_4$,
see Ref.~\cite{Paerschke2017} for details.

\subsection{Ladder spectrum in the \textit{t-}\texorpdfstring{$J_z$}{Jz} model} 

Assuming that the ground state at half-filling can be described by a classical N\'eel state with spin excitations, one can describe the motion of the hole in the AF background by an effective Hamiltonian   which naturally follows from an anisotropic \textit{t-J} model~\cite{Kane1989} by assuming a finite ratio of the exchange parameters, $\alpha = J_\perp / J_z$,
\begin{flalign}
\label{anisotropictJmodel}
\mathcal{H} = \sum_{\textbf{q}}  \omega_{\textbf{q}} a_{\textbf{q}}^{\dagger} a_{\textbf{q}}+ \nonumber\\
\frac{zt}{\sqrt{N}} \sum_{\textbf{k},\textbf{q}} {a_{\textbf{q}}^{\dagger} h_{\textbf{k}}^{\dagger} h^{\phantom{\dagger}}_{\textbf{k} + \textbf{q}} \left( u_{\textbf{q}}\gamma_{\textbf{k-q}}+v_{\textbf{q}}\gamma_{\textbf{k}}\right)+ \mathrm{h.c.}},  
\end{flalign}

where a charge excitation is represented by a spinless fermion with creation operator $h_{\textbf{k}}^{\dagger}$ and spin excitations are represented by the boson operators $a_{\textbf{q}}^{\dagger}$.
The spin-wave dispersion $\omega_{\textbf{q}} = z s J\left(1-\delta\right)^{2} \nu_{\textbf{q}}$ where $s=1/2$ is the spin and $z$ is the coordination number of the underlying square lattice. 
The Bogoliubov factors are given by
\begin{flalign}
      \nu_{\textbf{q}} &= \sqrt{1-\alpha^{2} \gamma_{\textbf{q}}^{2}},\nonumber \\
      u_{\textbf{q}} &= \sqrt{\frac{1+\nu_{\textbf{q}}}{2\nu_{\textbf{q}}}},\\
      v_{\textbf{q}} &= -\mathrm{sign}(\gamma_{\textbf{q}})\sqrt{\frac{1-\nu_{\textbf{q}}}{2\nu_{\textbf{q}}}}.\nonumber    
       \end{flalign}
The coupling of the hole to magnons is described by $\gamma_{\textbf{q}} =\frac{1}{z}\sum\limits_{\vec{\tau}}{cos \textbf{q} \cdot \vec{\tau}}$. 
Substituting $\alpha = 0$ to Eq.~(\ref{anisotropictJmodel}) we obtain the polaron representation of the \textit{t-J}$_{z}$ model,

\begin{flalign}
\label{tJzmodel}
\mathcal{H} = \omega \sum_{\textbf{q}} a_{\textbf{q}}^{\dagger} a_{\textbf{q}} + \frac{zt}{\sqrt{N}} \sum_{\textbf{k},\textbf{q}} \gamma_{\textbf{k}} 
a_{\textbf{q}}^{\dagger} h_{\textbf{k}}^{\dagger} h^{\phantom{\dagger}}_{\textbf{k} + \textbf{q}}+\mathrm{h.c.},
\end{flalign}
where the coefficients become $\textbf{q}$-independent: $\omega = szJ_{z}$ and $\gamma_{\textbf{k}} = \frac{1}{2}(\cos k_{x} + \cos k_{y})$.

To show that indeed different excitations in the ladder spectrum of the \textit{t-J}$_{z}$ model can be directly observed in the tunneling spectroscopy experiment, 
we map the polaronic Hamiltonian~(\ref{anisotropictJmodel}) onto an effective system of free fermions and bosons
\begin{flalign}
\label{freeel}
\tilde{\mathcal{H}} =  \sum_{\textbf{q}} \tilde{\omega}_{\textbf{q}}\alpha_{\textbf{q}}^{\dagger} \alpha_{\textbf{q}} + \sum_{\textbf{k}} \tilde{\varepsilon}_{\textbf{k}}  f_{\textbf{k}}^{\dagger} f^{}_{\textbf{k}},
\end{flalign}
where the new effective Hamiltonian $\tilde{\mathcal{H}}$ is related to the original Hamiltonian via a unitary transformation, 
\begin{eqnarray}
\label{UT}
\tilde{\mathcal{H}} = e^{X} \mathcal{H} e^{-X}.
\end{eqnarray}
Such a method enables to study the polaron excitations as projected on the effective free particle, which can directly couple to the tunneling electrons in an STS experiment.

The unitary transformation \eqref{UT} in general renormalizes the spin excitations of the background (first term of Eq.~(\ref{freeel})) 
and generates the second term of Eq.~(\ref{freeel}). Since the Eq.~(\ref{freeel}) has the quadratic diagonal form and the transformation \eqref{UT} is unitary, 
the energy quantities $\tilde{\varepsilon}_{\textbf{k}}$ and $\tilde{\omega}_{\textbf{q}}$ can be seen as eigenenergies of the original model. The transformation \eqref{UT}
can be constructed and numerically carried out by using the projective renormalization method (PRM) (see Ref.~\cite{Cho2016}). Within this method the polaronic term of the Hamiltonian
is integrated out in steps (1500 in the actual calculation), leading to the renormalization of the fermion and boson energy parameters. 

Using the unitary transformation \eqref{UT}, the one-particle spectral function can be calculated immediately,
\begin{flalign}
\label{PRM}
A_{\textbf{k}}(E) = |\tilde{a}^{0}_{\textbf{k}}|^2 \delta(E - \tilde{\varepsilon}_{\textbf{k}}) + \nonumber\\
\frac{1}{N} \sum_{\textbf{q}} |\tilde{a}^{1}_{\textbf{k},\textbf{q}}|^{2} 
\delta(E - \tilde{\varepsilon}_{\textbf{k}-\textbf{q}} - \tilde{\omega}_{\textbf{q}}) + \dots,
\end{flalign}
where $\tilde{a}^{0}_{\textbf{k}}$ and $\tilde{a}^{1}_{\textbf{k},\textbf{q}}$ are calculated in the renormalization process described above and represent the spectral weight of the
particular polaron excitation.

The internal excitations of the polaron can also be visualized by its wave function. Following~\cite{Reiter1994}, we write it in the form
\begin{flalign}
\label{Reiter}
|\Psi_{\textbf{k}} \rangle = a^{0}(\textbf{k}) f_{\textbf{k}}^{\dagger} |0 \rangle + \frac{1}{\sqrt{N}} \sum_{\textbf{q}} a^{1}(\textbf{k},\textbf{q}) f_{\textbf{k}-\textbf{q}}^{\dagger} 
\alpha_{\textbf{q}}^{\dagger} |0 \rangle + \dots.
\end{flalign}
Here, $|0 \rangle$ is the product of the hole vacuum and the spin-wave vacuum, and $\tilde{a}^{0}_{\textbf{k}}$, $\tilde{a}^{1}_{\textbf{k},\textbf{q}}$ are so-called Reiter coefficients.
We see that the wave function of the doped hole
can in principle be approximated by a superposition of the wave function of a free hole and $m$ wave functions of the hole dressed with $m$ magnons.

Fig.~\ref{fig:s3pol} shows the calculated effective dispersion of the spin polaron for different values of the ratio $\alpha = J_{\perp} / J_z$.
For small values of $\alpha$, i.e. close to the Ising limit, one clearly sees that the energy of the polaron increases as a function of its momentum $\textbf{k}$
in stair-step fashion: as soon as the momentum $\textbf{k}$ is sufficiently
large to produce a magnon, the polaron is raised to the next excited level. The ground state of the polaron is characterized by momentum states around $\textbf{k} = 0$ where only a finite range
of momentum values is occupied.
Since the values of $\alpha$ examined in Fig.~\ref{fig:s3pol} are small ($\alpha<0.25$), the magnon dispersion $\tilde{\omega}_{\textbf{q}}$ is almost momentum-independent and approximately equal 
to the magnon dispersion in the \textit{t-J}${_z}$ model (\ref{tJzmodel}).
For smaller values of $\alpha=0.028$ (shown in Fig.~\ref{fig:s3pol} with red circles), the polaron dispersion within each rung is quite flat,
whereas for larger values of $\alpha=0.25$ (green circles), the polaron becomes more dispersive. Overall, the polaron becomes less localized with the ratio $t/\omega$ decreasing.

The possibility to map the spin polaron model to an effective model of free charge carriers (dressed with characteristic ladder-like quasiparticle dispersion) indicates that
it must indeed be possible to detect internal excitations of spin polarons in an STM experiment.  

To get a better understanding of the nature of the polaron states shown on the Fig.1,
we calculate the first two Reiter coefficients $a^{0}(\textbf{k})$ and $a^{1}(\textbf{k},\textbf{q})$ from Eq.~(\ref{Reiter}) using perturbation theory with respect to the parameter $t/\omega$
assuming $\omega \gg t$ (strong coupling limit):
\begin{flalign}
& a^{0}(\textbf{k}) = 1 - \frac{1}{N} \sum_{\textbf{q}} \frac{\gamma_{\textbf{k}-\textbf{q}}^{2}}{\omega^{2}} \\
& a^{1}(\textbf{k},\textbf{q}) = \frac{\gamma_{\textbf{k}-\textbf{q}}}{\omega}.\nonumber
\end{flalign}
In this approximation, the spectral function of the hole has the form (similar to Eq.(~\ref{PRM}))
\begin{flalign}
A_{\textbf{k}}(E) = [a^{0}(\textbf{k})]^{2} \delta(E) + \frac{1}{N} \sum_{\textbf{q}} \frac{\gamma_{\textbf{k}-\textbf{q}}^{2}}{\omega^{2}} \delta(E - \omega).
\end{flalign}
This equation includes two different types of internal excitations. As one can see from the momentum dependence of the Reiter coefficients, the 
lowest excitation $a^{0}(\textbf{k})$ has $s$-wave character and represents a rather localized state of the hole. 
The second excitation $a^{1}(\textbf{k},\textbf{q})$ is spatially more extended due to its proportionality to $\cos$-functions, and the
sign of the coefficient changes as a function of momentum, which means that it is orthogonal to the first term.

\subsection{Relevance of the \textit{t-}\texorpdfstring{$J_z$}{Jz} ladder physics to Sr$_2$IrO$_4$}

\begin{figure}[!t]
\begin{center}
\includegraphics[width=\columnwidth]{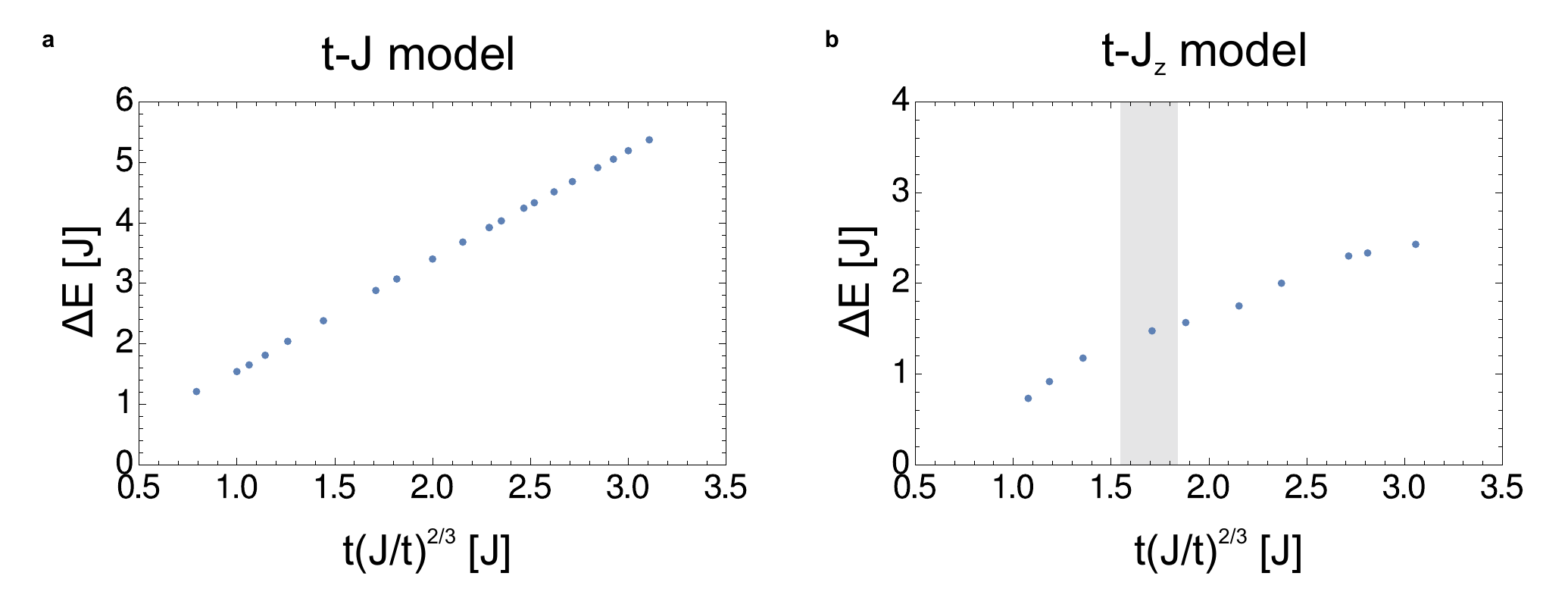}
\end{center}
 \caption{\label{fig:tJ_tJz}(a,b) 
Comparison of energy spacing between first and second excitation state of polaron scaling as a function of $J/t$ ratio: (A) \textit{t-}$J_z$ model, (B) material-specific \textit{t-J} model defined by
${\cal H}^+$ in Eq. \eqref{Ham_full_STS_methods}. In light gray the region of the $J/t$ values relevant for the Sr$_2$IrO$_4$ is shown: $J=0.06$ eV, first neighbor hoppings takes values from $0.224$ eV to
$0.373$ eV depending on the orbital character.}
\end{figure}

\begin{figure}[!t]
\begin{center}
\includegraphics[width=\columnwidth]{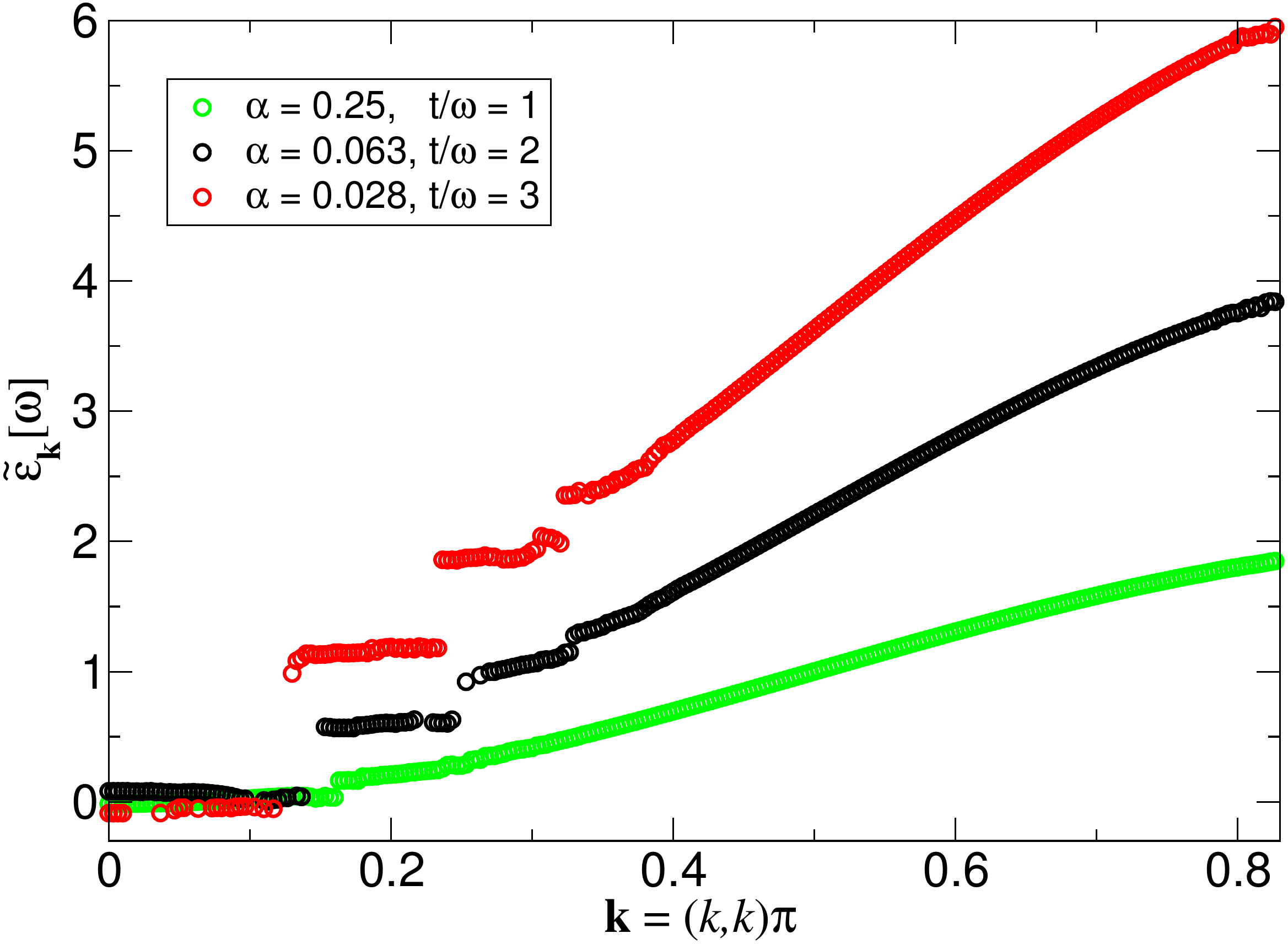}
\end{center}
 \caption{\label{fig:s3pol} 
Polaronic quasiparticle dispersion for the effective Hamiltonian of the anisotropic \textit{t-J} model given by Eq. \eqref{anisotropictJmodel} calculated for three different values of 
the ratio  $\alpha = J_\perp / J_z$. The value of $t$ is fixed throughout the calculation. The spectrum becomes more ladder-like as $\alpha$ approaches the Ising limit $\alpha = 0$. For the
lowest value of $\alpha$ (red circles) the value of $t/\omega$ lies in the relevant for Sr$_2$IrO$_4$ parameter region (as indicated in Fig.~\ref{fig:tJ_tJz}(b)) and the energy 
spacing between the first and second excitation is of the order of $J$.}
\end{figure}

To show the relevance of the above discussed theory to the case of Sr$_2$IrO$_4$ we have calculated the scaling of the energy spacing between first and second excited states on the positive side of the tunneling
conductance as a function of $J/t$ ratio for the material specific model \eqref{Ham_full_STS_methods}. It is known~\cite{Bulaevskii1968} that for the \textit{t-}$J_z$ model this energy spacing scales as 
$t(J_z/t)^{2/3}$, see fig. \ref{fig:tJ_tJz}(a). As one can see on the Fig. \ref{fig:tJ_tJz}(b), the energy gap calculated for model \eqref{Ham_full_STS_methods} follows the same law in the region of parameters relevant
to the real material (shown in light gray).

\subsection{Spin-polaron ladder spectrum signatures in STS}
In order to demonstrate the representative character of the $dI/dU$ spectrum shown in Fig. 2 (b), different spectra are shown in this section. These spectra were taken at different points in different topographies. For guidance the spin-polaron features present in the data are marked with arrows as shown in Figs. 2 (b).

\begin{figure*}[!t]
\begin{center}
\includegraphics[width=\linewidth]{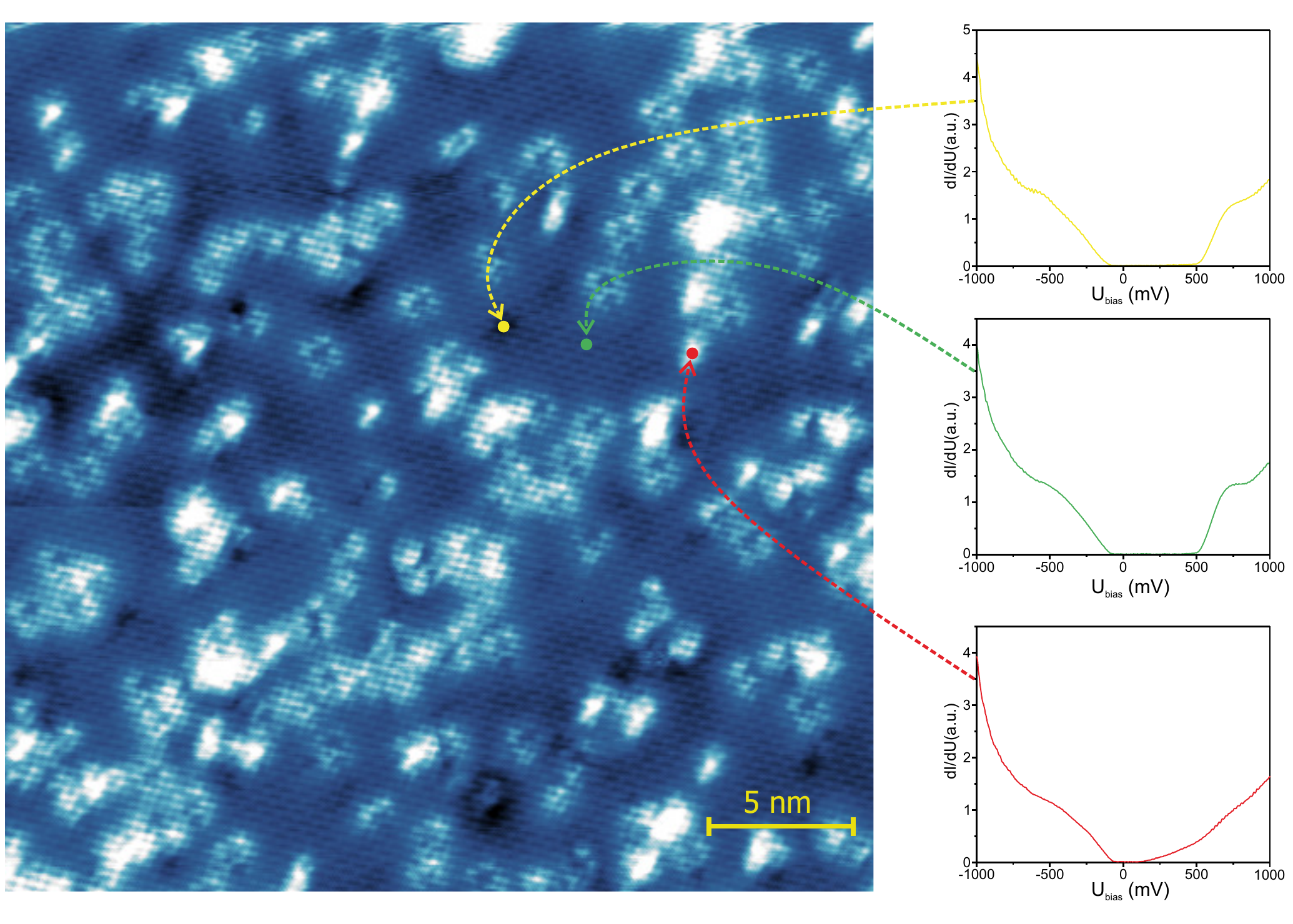}
\end{center}
\caption{\label{fig:si1} Topography image of  $20.5$ nm $\times\, 20.5$ nm measured at  $T=8.9$ K with $I_{Tunnel} = 200$ pA and $ U_{bias}= 1$ V, with the spectrum uses in FIG. 2 (b)  taken at a clean area of the atomically resolve surface. The $dI/dU$ are measured with an external lock-in amplifier with $f_{mod}=\, 1.111 $ kHz and $U_{mod}= \, 20$ mV (rms). }
\end{figure*}

In Fig. ~\ref{fig:si1}, three spectra taken at the atomically resolved surface are shown. The spectra were taken at different locations of the surface, darker area (yellow curve), clean area (green curve) and on top of a frequent defect (red curve). The features of the spin-polaron spectrum spectra can be distinguished in the clean and dark area. 
In both cases the characteristic ladder-like signature is present on the positive bias side. On top of the defect spectral intensity is found inside the gap, in agreement with previous high temperature findings \cite{yan2015electron}.

\begin{figure*}[!t]
\begin{center}
\includegraphics[width=\linewidth]{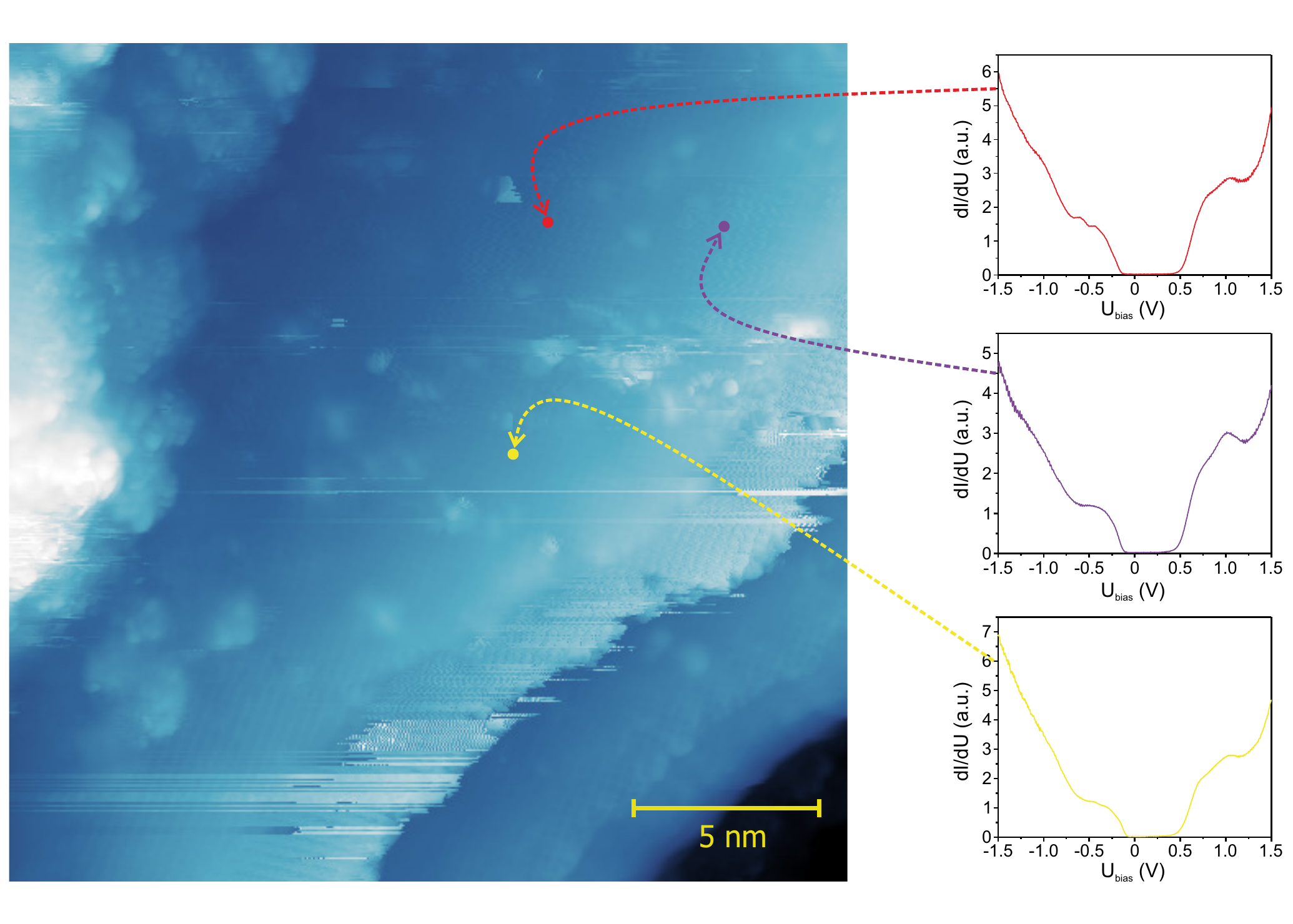}
\end{center}
\caption{ 
\label{fig:si2}
 Topography image of $22.5$ nm $\times \, 22.5$ nm showing a terrace of the crystal, measured at  $T=8.9$ K with $I_{Tunnel} = 200$ pA and $ U_{bias}= 900$ mV,  with spectra taken at different locations. All the $dI/dU$ are measured with large Bias values ($-1.5$ V to $1.5$ V) with an external lock-in amplifier with $f_{mod}=\, 1.111$ kHz and $U_{mod}= \, 20 $ mV (rms). }

\end{figure*}

In Fig. ~\ref{fig:si2} a different topography location is shown. In this case the topography presents a terrace between steps in the cleaved surface of Sr$_{2}$Ir$_{4}$. Even without the best experimental conditions, the $dI/dU$ measured with large bias values ($-1.5$ V to $1.5$ V) probe that the features of the spin polaron are still present.

\begin{figure*}[!t]
\begin{center}
\includegraphics[width=\linewidth]{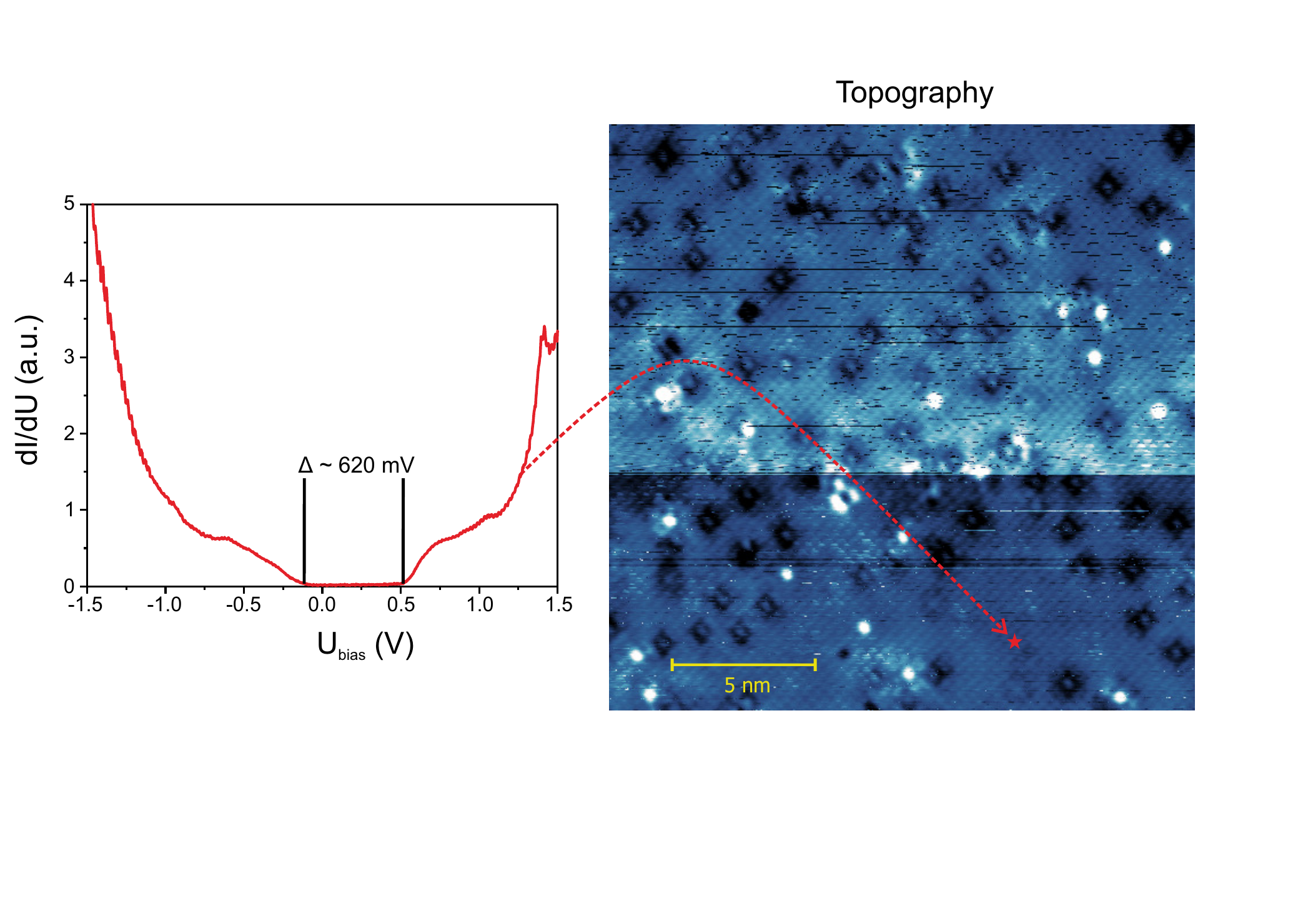}
\end{center}
\caption{\label{fig:si3} Topography image of  $20.5$ nm $\times\, 20.5$ nm measured at  $T=8.9$ K with $I_{Tunnel} = 200$ pA and $ U_{bias}= 1$ V, with a spectra taken at a clean area in the atomically resolve surface. The $dI/dU$ are measured with an external lock-in amplifier with $f_{mod}=\, 1.111 $ kHz and $U_{mod}= \, 20$ mV (rms). }
\end{figure*}

In Fig. ~\ref{fig:si3} an atomically resolved surface (measured with some experimental noise) is plotted for the same location as the $dI/dU$ spectrum shown in Fig. 2 (b). We show three additional spectra referring to the darker area (yellow curve), and clean area (green and red curve). The features of the ladder spectra can be clearly distinguished in the clean area.

In general it can be noted that the signatures of the confined spin polaron are more distinguishable at the positive bias. In the negative side the dispersive character and internal degrees of freedom of the charge excitation make it harder to recognize them. 

\begin{figure*}[!t]
\begin{center}
\includegraphics[width=\linewidth]{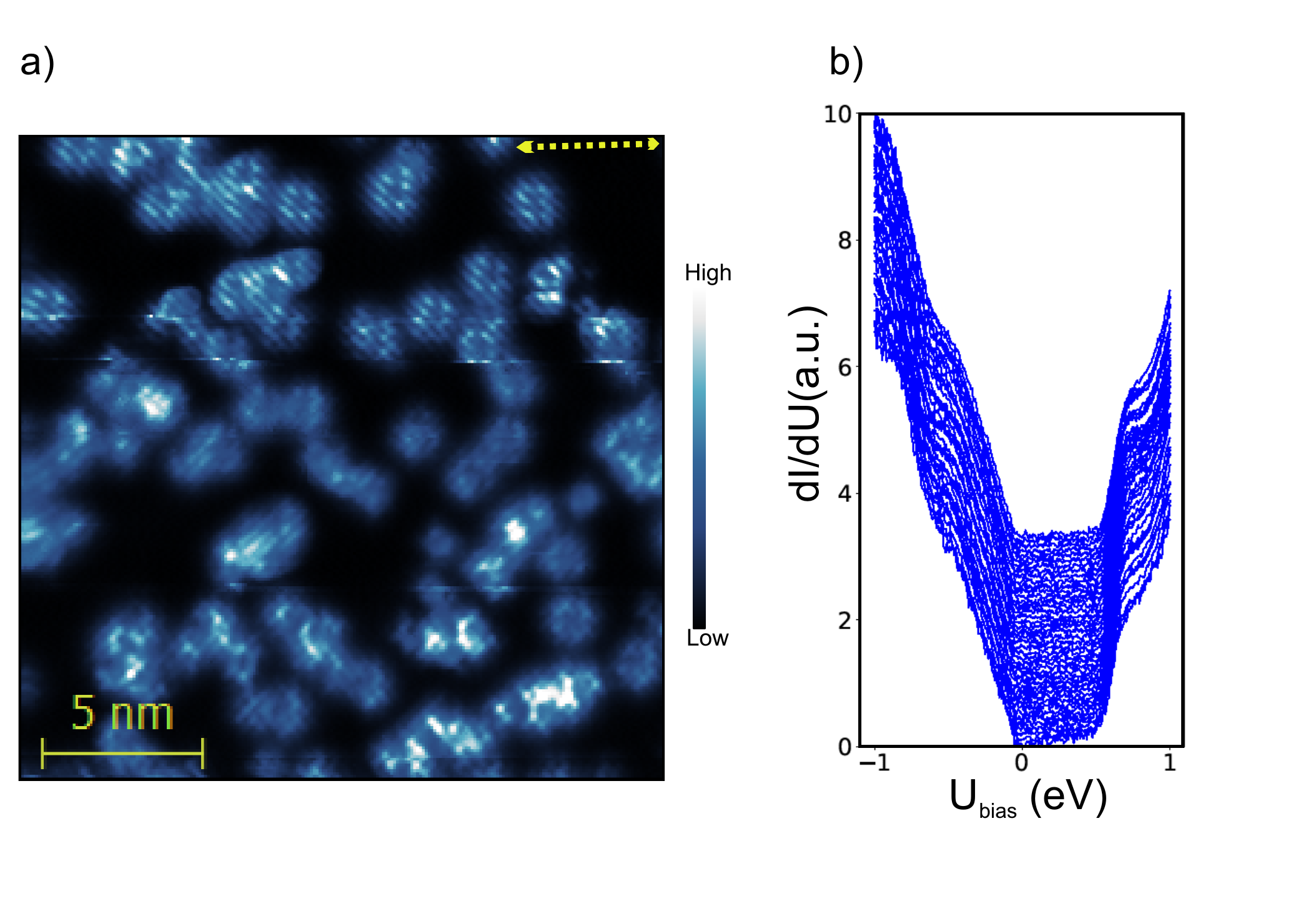}
\end{center}
\caption{\label{fig:si5} (a)SI-STM map at 303.7 meV. Measured at $T=9.3 $ K with $I_{Tunnel} = 300$ pA and $ U_{bias}= 1.0$ V, the  $dI/dU$ spectra were measured with an external lock-in amplifier with $f_{mod}=\, 1.111$ kHz and $U_{mod}= \, 15 $ mV (rms) for 271 energy point from $-1.0$ V to $1.0$ V in an area of $20$ nm $\times\, 20$ nm with a $224$  $\times\, 224$ grid.
(b)  Local density of states spectra along the the yelow line in (a). }

\end{figure*}

In Fig. ~\ref{fig:si5} an spectroscopy-imaging STM (SI-STM) data is shown in (a) at 300 meV (inside the gap) where the deffects can be atomically resolved. In (b) we plot the local density of states along the yelow line over a clean area in (a). In all spectra the first spin-polaron peak as well as the beggining of the second peak are clearly recognizable. Thisis a proof of the universality of the spin-polaron in the clean areas of the sample.

\subsection{Determination of the peaks positions}
\begin{figure*}[!t]
\begin{center}
\includegraphics[width=\linewidth]{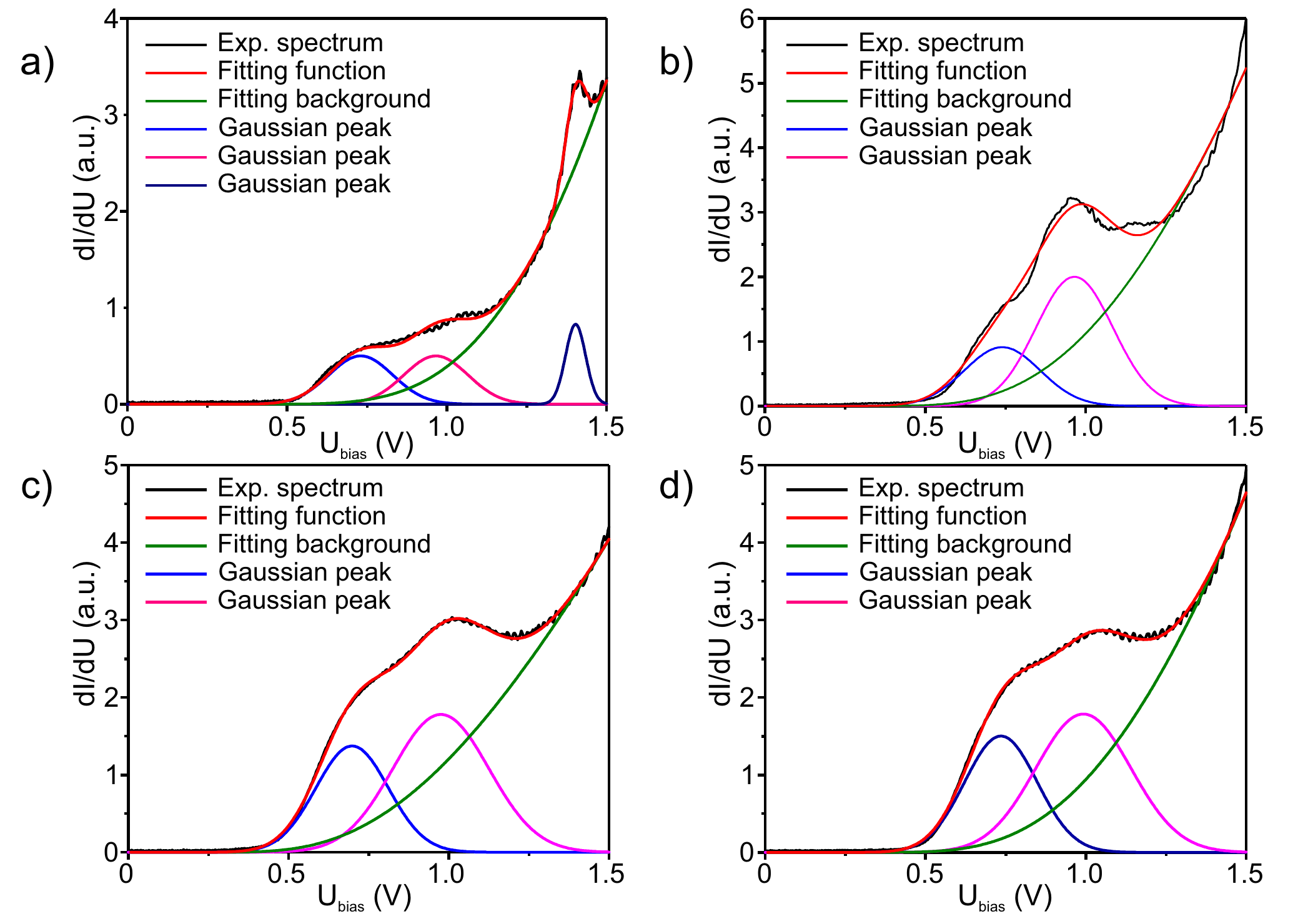}
\end{center}
\caption{\label{fig:si6}Fitting procedure to find the position of the spin-polaron peaks of the spectra (a) from Fig.~\ref{fig:si2} and (b),(c) and (d) from Fig.~\ref{fig:si3}. }
\end{figure*}

In order to determine the position of the peaks corresponding to the spin polaron we use a fitting function $F(x) = F_B(x) + \sum_i F_i(x)$ for the measured spectrum which consists of a bosonic background
\begin{linenomath*}
\begin{align}
	    \label{fit_fucn}
F_{B}(x)= \frac{a}{(e^{b/x}-1)} 
\end{align}
\end{linenomath*}
and Gauss functions for the intensity of the peaks, 
\begin{linenomath*}
\begin{align}
	    \label{gaussian}
F_i(x) = A_{i} \exp \left (\frac{-(x-\bar{x}_{i})^{2}}{2 \sigma_{i}^{2}}\right),
\end{align}
\end{linenomath*}
where $A_{i}$,  $\bar{x}_{i}$, and $\sigma_{i}$ are independent parameters for each peak. The position of the spin-polaron peaks is given by  $\bar{x}_{i}$.
The resulting plots can be seen in Fig.~\ref{fig:si6}

The function~(\ref{fit_fucn}) is a phenomenological
description of a bosonic background which 
is introduced to simulate the dissipation of the energy 
which is transferred to the system through the
tunneling current. Thus, under the assumption that the 
background is fully described by a system of bosons 
the background contribution to the tunneling response 
takes the form of a boson distribution function where 
the temperature plays the role of dissipated energy which
is set proportional to the bias voltage. In addition to this background the intrinsic excitations of the spin polaron are fitted by Gauss functions where the 
corresponding positions, amplitudes, and widths are 
extracted.   

For the spectrum in Fig.~\ref{fig:si6}(a) we find three gaussian peaks with the fitting parameters given by the following table: \\

\begin{tabular}{|l|l|l| }
\hline
\multicolumn{3}{ |c| }{Spectrum (a)} \\
\hline

\multirow{2}{*}{Background} & a & 277.594 (a.u.)\\ 
 & b & 6.416 (meV)\\ \hline
\multirow{3}{*}{Gaussian peak 1} & Mean &0.730  (meV) \\
 &$ \sigma$ & 0.100 (meV)\\
 & Amplitude & 0.500 (a.u.)\\ \hline
\multirow{3}{*}{Gaussian peak 2} & Mean & 0.966 (meV)\\
 & $\sigma $& 0.100 (meV)\\
 & Amplitude & 0.500 (a.u.)\\ \hline

 \multirow{3}{*}{Gaussian peak 3} & Mean &1.403 (meV)\\
 & $\sigma $& 0.031 (meV)\\
 & Amplitude & 0.828 (a.u.)\\  \hline
 
\end{tabular}

The spectra (b), (c) and (d) in Fig. ~\ref{fig:si6} were taken under the same tunneling conditions. They do no present the higher energy peak present in the spectrum (a) at 1.4 eV. Since this peak is absent we use two gaussian peaks for the spin-polaron ground state and first excitation in the fitting procedure. Fitting parameters:

 \begin{tabular}{|l|l|l| }
\hline
\multicolumn{3}{ |c| }{Spectrum (b)} \\
\hline
\multirow{2}{*}{Background} & a & 100.0 (a.u.)\\ 
 & b & 4.502 (meV)\\ \hline
\multirow{3}{*}{Gaussian peak 1} & Mean &0.740 (meV)\\
 &$ \sigma$ & 0.118 (meV)\\
 & Amplitude & 0.910 (a.u.)\\ \hline
\multirow{3}{*}{Gaussian peak 2} & Mean & 0.965 (meV)\\
 & $\sigma $& 0.120 (meV)\\
 & Amplitude & 2.000 (a.u.)\\ \hline

\end{tabular}

 \begin{tabular}{|l|l|l| }
\hline
\multicolumn{3}{ |c| }{Spectrum (c)} \\
\hline
\multirow{2}{*}{Background} & a & 35.058 (a.u.)\\ 
 & b & 3.4 (meV)\\ \hline
\multirow{3}{*}{Gaussian peak 1} & Mean &0.698 (meV)\\
 &$ \sigma$ & 0.113 (meV)\\
 & Amplitude & 1.373 (a.u.)\\ \hline
\multirow{3}{*}{Gaussian peak 2} & Mean & 0.975 (meV)\\
 & $\sigma $& 0.150 (meV)\\
 & Amplitude & 1.780 (a.u.)\\ \hline

\end{tabular}

 \begin{tabular}{|l|l|l| }
\hline
\multicolumn{3}{ |c| }{Spectrum (d)} \\
\hline
\multirow{2}{*}{Background} & a & 100.0 (a.u.)\\ 
 & b & 4.675 (meV)\\ \hline
\multirow{3}{*}{Gaussian peak 1} & Mean &0.735 (meV)\\
 &$ \sigma$ & 0.113 (meV)\\
 & Amplitude & 1.502 (a.u.)\\ \hline
\multirow{3}{*}{Gaussian peak 2} & Mean & 0.992 (meV)\\
 & $\sigma $& 0.149 (meV)\\
 & Amplitude & 1.786 (a.u.)\\ \hline

\end{tabular}

The background parameter $a$ is the amplitude of the bosonic background and $b$ has the role of the bosonic energy in the system. The similar values of $b$ for the different spectra indicate a uniform applicability of our used background fitting function  \eqref{fit_fucn}. Further consistency of our fitting method is shown by very similar positions of the same peak in different spectra. The slight variation of the peak width within the same spectrum could be caused by many-body phenomena.

\clearpage
\newpage

\clearpage
\newpage

\end{document}